# An Econophysical dynamical approach of expenditure and income distribution in the UK

**Elvis Oltean, Fedor V. Kusmartsev**

e-mail: elvis.oltean@alumni.lboro.ac.uk

**Abstract:** We extend the exploration regarding dynamic approach of macroeconomic variables by tackling systematically expenditure using Statistical Physics models (for the first time to the best of our knowledge). Also, using polynomial distribution which characterizes the behavior of dynamical systems in certain situations, we extend also our analysis to mean income data from the UK that span for a time interval of 35 years. We find that most of the values for coefficient of determination obtained from fitting the data from consecutive years analysis to be above 80%. We used for our analysis first degree polynomial, but higher degree polynomials and longer time intervals between the years considered can dramatically increase goodness of the fit. As this methodology was applied successfully to income and wealth, we can conclude that macroeconomic systems can be treated similarly to dynamic systems from Physics. Subsequently, the analysis could be extended to other macroeconomic indicators.

## 1. Introduction

Most of the papers analyze macroeconomic variables such as income and wealth distribution using time intervals of one year. The dynamic evolution of the macroeconomic systems is done by using the temporal evolution of the coefficients obtained from the distribution fitting the annual data and by displaying their temporal values.

We proposed a different approach of income, wealth, or expenditure distribution by employing a distribution which uses the difference between the values calculated for deciles of income from two different years (regardless if they are consecutive or not) instead of the annual values [1]. The distribution that best describes such phenomena is the polynomial distribution using normal values for the axes instead of logarithmic values (log-log scales). This was a very important finding given that polynomials are used in dynamic systems to describe their behavior in particular situations. Thus, this may point to a more strict correlation between dynamic systems from Physics and macroeconomic systems. The results obtained from applying this method proved its applicability, given that coefficient of determination for the statistical distribution used was in most of the cases above 80 %. It is noteworthy that most of the analyses were performed on consecutive years. This paper extends the dynamic analysis to other income data and tackles systematically expenditure data, which were provided extensively only by the UK.

A first among the analyses of all sorts of macroeconomic variables was expenditure or consumption analysis. This paper explores systematically for the first time the dynamics of expenditure and, therefore, we believe to be very useful in the analysis of living standards and inequality. Thus, expenditure is more relevant, in our opinion, than income and wealth as it shows the amount of goods and services purchased.

Another novelty introduced by this article is the method used to calculate different values for the deciles. So far, the analyses used mean income and upper limit on income. The national office of statistics from the UK uses another term in order to describe values for deciles of population. Lower bound is taken into account in order to analyse the activity of all 10 deciles such that the values for the tenth decile (which includes the upper income segment of population) can be used. Lower bound values are the opposite of upper limit on income values (term coined by national statistical body of Finland [2]), and we shall nominate it as lower limit on income.

## 2. Short Literature Review and Theoretical Framework

In the analysis of income and wealth, the most used were deciles of population ranked in increasing order of the values for income and/or wealth, calculated for time intervals of one year. The method used for analysis was mean (average) income, calculated as the sum divided to the number of persons included in that decile. The most utilized were Boltzmann-Gibbs, Bose-Einstein, and lognormal (Gibrat) distributions. In addition, the most explored type of income in the analyses was disposable income, other types such as gross income being considered very rarely [3-8].

A milestone for this area is the work of Clementi and Gallegati [9], which is the most interesting contribution to the dynamic evolution of macroeconomic variables. The authors consider the evolution of real mean income per capita and GDP growth (economic cycle) in the Italian economy by using biennial data from the time interval 1987-2002. The main findings of this paper are that lognormal distribution describes low and middle income distribution, while Pareto describes the upper segment of income of population. However, the slope of Pareto distribution and the curvature of the lognormal distribution are different from one year to another. Moreover, another finding is about the fact that personal income and GDP growth are well fit by a Laplace distribution. Thus, the evolutions of personal income and GDP for the Italian economy are similar, which may lead to a possible common mechanism that would characterize the similarity of growth dynamics for both indicators.

Polynomial distribution was first used in order to describe the static evolution of income distribution for annual values [10]. The results showed that polynomial distribution fits very well the data, having values for coefficient of determination higher than 95% in all cases.

The method proposed [1] has the disadvantage that the analyses which used it were performed only on consecutive years. Therefore, negative differences were caused by drops in the value for the same decile from one year to the next. Subsequently, the shape of the distribution was affected when this occurred. It is noteworthy that when the analyses performed on the first and last years (from any data set we used) showed values for coefficient of determination to be among the highest values. This is explainable on the long run, especially in the case of the nominal income, as deciles have increasing values over the time. Thus, the probability to get negative differences is lower. However, such a possibility is not entirely excluded given the cyclical evolution of an economy, since this affects especially the real income.

## 3. Methodology

We use this methodology as polynomials model the behavior of dynamic systems in certain situations. Thus, we can relate the behavior of macroeconomic systems to the ones from Physics. Therefore, we can treat macroeconomic systems as being dynamic, including the values of its indicators and variables.

Most authors in the field of income and wealth distribution used nominal values. It is known that real income is a better notion in the analysis of income. This is because nominal income figures may be misleading, while real income takes into account the inflation which may diminish the increase of money quantity that a household or a person earns [11]. Having that in mind, we believe that using the real income data where these were available was desirable in the exploration of income and expenditure (about their evolution in time). Moreover, given that our paper focuses on dynamic aspects of income and expenditure, the usage of real values for income and expenditure is more appropriate.

The deciles of expenditure are calculated based on income deciles. Thus, for each decile the expenditure of the population contained in that income category is summed up and divided to the number of persons (in the case of mean expenditure). Also, the lowest bound regarding expenditure from a household whose income falls into that decile of income is considered to be the lower limit on expenditure for the respective decile.

The methodology we used in order to fit the data is the cumulative method. Thus, on the x axis we represented the income or expenditure difference, calculated for each decile by taking into account annual values. The difference may be negative or positive and is ranked in increasing order and afterwards summed up. On the y-axis, we displayed cumulated probability density of the population which has a certain increase or decrease of income or expenditure and which is strictly higher than a certain threshold chosen from x-axis. Consequently, both for mean (average) and lower limit on income/expenditure we will display ten values. Unlike in the case of static approach, where everyone was considered to have some sort of income and, therefore, for income equal to zero the population probability was 100%, the situation is a bit different. The income and expenditure differences could be negative, so it is possible to get that given decreases in income (hence negative values for decile) the cumulated probability assigned is no longer 100% for zero expenditure and/or income. The graphics will use normal values, not logarithmic ones (log-log scale).

More formally, the methodology we use is as follows: let $x_{11}, x_{12}, \ldots\ldots\ldots x_{10}$ be the values assigned for each decile in the first year and let $x_{21}, x_{22}, \ldots\ldots\ldots X_{20}$ the values assigned for each decile in the second year analysed. Please note that the second year is chronologically following the first year (regardless is consecutive or not).

Let $\Delta x_1, \Delta x_2, \ldots\ldots\ldots \Delta x_n$ be such that $\Delta x_1 = x_{21} - x_{11}$ for the first decile, $\Delta x_2 = x_{22} - x_{21}$ for the second decile, and $\Delta x_{10} = x_{20} - x_{10}$; for the tenth second decile, where $\Delta x_1, \Delta x_2, \ldots\ldots\ldots \Delta X_{10}$ are real numbers (negative or positive).

Let $S$ be a set such that $S = \{\Delta x_1, \Delta x_2, \ldots\ldots\ldots \Delta X_{10}\}$. Let's assume that $x_A, x_B, x_C, x_D, x_E, x_F, x_G, x_H, x_I, x_J$ are such that

$x_{x_A} S, S, x_B$  $S, x_C S, x_D S, x_E S, x_F S, x_G S, x_H S,$

xJ∈S and xA<xB<xC<xD<xE<xF<xG<xH<xI<xJ

In the case of mean (average) values, for the lowest value in the decile difference the corresponding cumulated population probability is 90%, as the rest of the population (which belongs to the other nine deciles) has a higher value for increase (or decrease) in income or expenditure. Similarly, we get the values for the rest of differences in decile values. Subsequently, the set of plots for mean (average) income or expenditure is M= {($x_A$,90%), ($x_B$,80%), ($x_C$,70%), ($x_D$,60%), ($x_E$,50%), ($x_F$,40%), ($x_G$,30%), ($x_H$,20%), ($x_I$,10%), ($x_J$,0%)}.

In the case of lower limit on income/expenditure, (i.e.) the lowest difference of two deciles, the corresponding percentage of cumulated population density is 100%. Similarly, we continue with the rest of the values for the rest of the nine deciles. The graphical plot set is L={($x_A$,100), ($x_B$,90), ($x_C$,80), ($x_D$,70), ($x_E$,60), ($x_F$,50), ($x_G$,40), ($x_H$,30), ($x_I$,20), ($x_J$, 10)}.

The data used for analysis [12] were about households. In the case of income, the data were about annual values, while for expenditure the data were expressed in weekly values. All data used were about the UK and we considered them valuable especially considering their reliability and time interval of 35 years they span.

## 4. Results

The values for coefficients obtained from fitting the data are displayed in the Appendixes 1-5. Most of the values for the coefficient of determination are above 80%. The distributions fitting the data exhibit values for coefficient of determination lower than 80% in the harshest period of the crisis (2007-2010).

The data from outside the crisis interval exhibited a uniform shape given that all deciles showed more or less a uniform increase. Subsequently, the shape of the distribution was very similar the static approach ones, which fit the income annual values characterized by higher values for $R^2$. During crises, values for deciles regarding income and expenditure do not preserve the same shape for distribution, as some deciles have income and expenditure values which decrease, while the rest have a lower increase than before and, exceptionally, for some deciles the income and expenditure may increase even more. Moreover, in crises intervals the slope of the polynomial is mostly positive, unlike most distributions which have negative slope.

The goodness of the fit in the analyses can be increased by using higher degree polynomials (at least second degree) and longer time intervals between the years considered. We found that using both these ways can increase substantially the goodness of the fit. Moreover, there is a trade-off up to a certain extent between the length of time interval and the degree of the polynomial used. Thus, the longer the time interval, the lower can be the degree of the polynomial. This is explainable by the fact that (especially for nominal values) income generally increases on the long run (and subsequently for wealth and expenditure). Drops in income for longer time intervals occur only in cases of severe economic recession.

Regarding the dynamic behavior of these types of data in the fitting process, we can conclude that income data were better fitted by polynomial distribution than the expenditure data. Mean expenditure and lower limit on expenditure data were equally well fitted by polynomial distribution. Also, there is no significant difference in the goodness of the fit between gross and disposable expenditure.

We present few typical graphical examples in the Figures 1, 2, and 3.

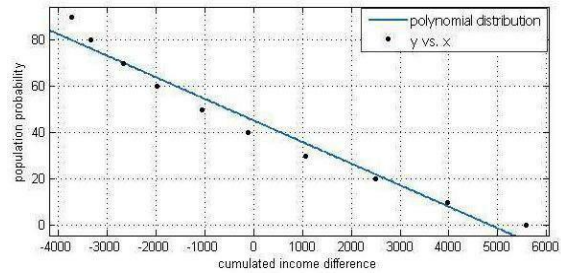

*Figure 1. Polynomial distribution applied to mean income difference for the time interval 2003-2002/2002-2001*

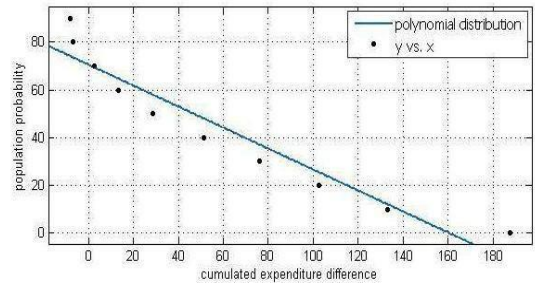

*Figure 2. Polynomial distribution applied to mean gross expenditure difference for the time interval 2010/2009*

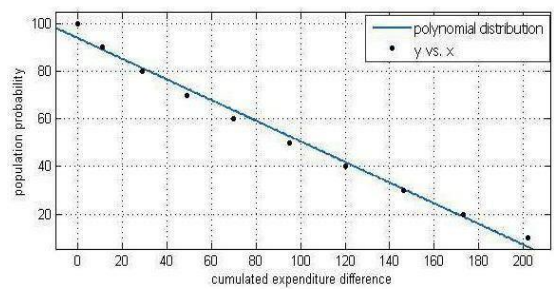

*Figure 3. Polynomial distribution applied to lower limit on disposable expenditure difference for the time interval 2011/2010*

## 5. Conclusions

We can conclude that the dynamic method using polynomial distribution is robust even using first degree polynomial. This methodology could be dramatically improved by using higher degree polynomials and longer time intervals.

This is the first time when we use for analysis lower limit on expenditure values for deciles. We consider it to be a very useful new way of calculating values for deciles used in analyses.

The expenditure can be described by the dynamic methodology proposed using polynomial distribution, given that data we used for our analysis are reliable.

Economic theory can benefit from this endeavor regarding correlation of income and expenditure. This could help in explaining how the taxes and social benefits affect (macroeconomic) aggregated consumption. More concretely, this could enable us to explore how taxes and social benefits influence living standard and inequality.

Further analyses should consider optimal use of this methodology with regard to the degree of polynomials and time interval having minimal number of parameters.

# Appendix

*Appendix 1. Coefficients from fitting polynomial distribution to mean income difference.*

| Year | P1 | P2 | $R^2$ (%) |
|---|---|---|---|
| 1978/1977 | 0.02381 | 132.1 | 88.95 |
| 1979/1978 | -0.0138 | 77.85 | 94.29 |
| 1980/1979 | -0.00928 | 80 | 94.34 |
| 1981/1980 | -0.0133 | 83.04 | 93.52 |
| 1982/1981 | -0.02679 | 76.18 | 92.02 |
| 1983/1982 | -0.0229 | 80.33 | 90.35 |
| 1984/1983 | -0.02002 | 74.69 | 92.65 |
| 1985/1984 | -0.01105 | 74.57 | 87.05 |
| 1986/1985 | -0.01251 | 69.3 | 83.61 |
| 1987/1986 | -0.009441 | 74.51 | 90.91 |
| 1988/1987 | -0.007041 | 69.1 | 85.18 |
| 1989/1988 | -0.01161 | 76.74 | 91.45 |
| 1990/1989 | -0.005722 | 72.87 | 85.18 |
| 1991/1990 | -0.009055 | 82.61 | 95.55 |
| 1992/1991 | -0.02581 | 67.39 | 97.72 |
| 1993/1992 | -0.01989 | 55.22 | 64.62 |
| 1995-1994/1993 | -0.01887 | 80.61 | 94.11 |
| 1996-1995/1995-1994 | -0.02059 | 75.05 | 94.18 |
| 1997-1996/1996-1995 | -0.00694 | 69.84 | 83.64 |
| 1998-1997/1997-1996 | -0.008682 | 74.44 | 90.5 |
| 1999-1988/1998-1997 | -0.01116 | 73.17 | 82.53 |
| 2000-1999/1999-1988 | -0.008217 | 65.95 | 81.94 |
| 2001-2000/2000-1999 | -0.008939 | 85.46 | 98 |
| 2002-2001/2001-2000 | -0.005186 | 71.46 | 84.6 |
| 2003-2002/2002-2001 | -0.009344 | 45.23 | 96.69 |
| 2004-2003/2003-2002 | -0.009791 | 45.97 | 53.54 |
| 2005-2004/2004-2003 | -0.007156 | 83.62 | 95.67 |
| 2006-2005/2005-2004 | -0.01071 | 59.57 | 77.43 |
| 2007-2006/2006-2005 | -0.006951 | 77 | 89.54 |
| 2008-2007/2007-2006 | -0.01479 | 47.85 | 74.9 |
| 2009-2008/2008-2007 | -0.01103 | 57.47 | 72.68 |
| 2010-2009/2009-2008 | -0.01062 | 71.92 | 89.63 |
| 2011-2010/2010-2009 | 0.008897 | 123.8 | 73.95 |
| 2012-2011/2011-2010 | -0.004618 | 75.31 | 90.64 |
| 2012-2011/1977 | -0.0003679 | 81.08 | 93.59 |

*Appendix 2. Coefficients from fitting polynomial distribution to mean disposable expenditure difference.*

| Year | P1 | P2 | $R^2$ (%) |
|---|---|---|---|
| 2002-2001/2000-2001 | -0.7087 | 73.58 | 89.26 |
| 2003-2002/2001-2000 | -1.042 | 71.72 | 88.04 |
| 2004-2003/2003-2002 | -0.6977 | 71.62 | 88.64 |
| 2005-2004/2004-2003 | -0.5392 | 81.6 | 96.96 |
| 2006-2005/2005-2004 | -0.6701 | 45.99 | 58.78 |
| 2006/2006-2005 | -0.4573 | 44.09 | 80.93 |
| 2007/2006 | -0.4964 | 10.08 | 53.02 |
| 2008/2007 | -0.5036 | 49.66 | 68.06 |
| 2009/2008 | 0.3494 | 104.4 | 34.53 |
| 2010/2009 | -0.43 | 67.27 | 87.22 |
| 2011/2010 | -0.6179 | 48.7 | 73.83 |
| 2012/2011 | -0.5696 | 25.16 | 53.9 |
| 2012/2000-2001 | -0.1065 | 88.76 | 97.95 |

*Appendix 3. Coefficients from fitting polynomial distribution to mean gross expenditure difference.*

| Year | P1 | P2 | $R^2$ (%) |
|---|---|---|---|
| 2002-2001/2000-2001 | -0.6124 | 64.74 | 85.36 |
| 2003-2002/2001-2000 | -0.9668 | 64.16 | 86.99 |
| 2004-2003/2003-2002 | -0.6708 | 70.44 | 88.2 |
| 2005-2004/2004-2003 | -0.559 | 82.22 | 95.4 |
| 2006-2005/2005-2004 | 0.1546 | 120 | 92.04 |
| 2006/2006-2005 | -0.1011 | 79.82 | 94.88 |
| 2007/2006 | -0.8521 | 27.14 | 51.58 |
| 2008/2007 | -0.5397 | 50.26 | 70.2 |
| 2009/2008 | 0.3987 | 110 | 47.31 |
| 2010/2009 | -0.4397 | 70.49 | 91.86 |
| 2011/2010 | -0.716 | 62.76 | 90.4 |
| 2012/2011 | -0.7118 | 24.24 | 58.96 |
| 2012/2000-2001 | -0.1069 | 89.61 | 98.4 |

*Appendix 4. Coefficients from fitting polynomial distribution to lower limit on disposable expenditure difference.*

| Year | P1 | P2 | $R^2$ (%) |
|---|---|---|---|
| 2002-2001/2000-2001 | -0.3293 | 83.32 | 89.74 |
| 2003-2002/2001-2000 | -0.6871 | 94.58 | 99.03 |
| 2004-2003/2003-2002 | -2.337 | 54.07 | 83.56 |
| 2005-2004/2004-2003 | -0.3823 | 86.35 | 91.62 |
| 2006-2005/2005-2004 | -0.779 | 65.36 | 73.13 |
| 2006/2006-2005 | -0.5135 | 88.43 | 95.26 |
| 2007/2006 | -0.6334 | 92.05 | 97.62 |
| 2008/2007 | -0.07416 | 76.22 | 80.96 |
| 2009/2008 | 0.1171 | 146.7 | 72.87 |
| 2010/2009 | -1.043 | 74.61 | 76.86 |
| 2011/2010 | -0.4326 | 93.72 | 98.97 |
| 2012/2011 | 5.673 | 144.1 | 25.79 |
| 2012/2000-2001 | -0.0745 | 89.87 | 95.95 |

*Appendix 5. Coefficients from fitting polynomial distribution to lower limit on gross expenditure difference.*

| Year | P1 | P2 | $R^2$ (%) |
|---|---|---|---|
| 2002-2001/2000-2001 | -0.2763 | 81.47 | 86.88 |
| 2003-2002/2001-2000 | -0.7086 | 94.61 | 99.16 |
| 2004-2003/2003-2002 | -1.198 | 86.38 | 94.36 |
| 2005-2004/2004-2003 | -0.3004 | 83.9 | 89.04 |
| 2006-2005/2005-2004 | -0.6509 | 64.89 | 72.19 |
| 2006/2006-2005 | -0.4234 | 87.26 | 94.27 |
| 2007/2006 | -0.4475 | 87.98 | 95.22 |
| 2008/2007 | -0.492 | 67.45 | 66.82 |
| 2009/2008 | 0.2688 | 74.65 | 2.2 |
| 2010/2009 | -1.172 | 74.22 | 71.87 |
| 2011/2010 | -0.3607 | 92.62 | 98.44 |
| 2012/2011 | 1.521 | 141 | 68.69 |
| 2012/2000-2001 | -0.06751 | 88.73 | 94.72 |

## References

[1] E. Oltean, F.V. Kusmartsev, A dynamic approach to income and wealth distribution, unpublished

[2] "Upper limit on income by decile group in 1987–2009 in Finland", URL: http://www.stat.fi/til/tjt/2009/tjt_2009_2011-0520_tau_005_en.html

[3] A. Dragulescu, V. M. Yakovenko, "Statistical Mechanics of Money, Income, and Wealth: A Short Survey", URL: http://www2.physics.umd.edu/~yakovenk/econophysics.html

[4] A. Dragulescu, V.M. Yakovenko, "Statistical mechanics of money", Eur. Phys. J. B 17, 723-729 (2000).

[5] A.Dragulescu, V.M. Yakovenko, "Evidence for the exponential distribution of income in the USA", Eur. Phys. J. B 20, 585-589 (2001)

[6] A. C. Silva, V. M. Yakovenko, "Temporal evolution of the "thermal" and "superthermal" income classes in the USA during 1983–2001", Europhys. Lett., 69 (2), pp. 304–310 (2005)

[7] K. E. Kurten and F. V. Kusmartsev, "Bose-Einstein distribution of money in a free-market economy". Physics Letter A Journal, EPL, 93 (2011) 28003

[8] F. V. Kusmartsev, "Statistical mechanics of economics", Physics Letters A 375 (2011) 966–973

[9] F. Clementi, M. Gallegati, "Power Law Tails in the Italian Personal Income Distribution", *Physica A*, volume 350, pages 427–438, 2005.

[10] E. Oltean, F.V. Kusmartsev, "A polynomial distribution applied to income and wealth distribution", Journal of Knowledge Management, Economics, and Information Technology, 2013, Bucharest, Romania

[11] URL:http://www.investopedia.com/terms/r/realincome.asp#axzz2J6oHpvFu

[12] URL:http://www.ons.gov.uk/ons/search/index.html?newquery=household+income+by+decile&newoffset=0&pageSize=50&sortBy=&applyFilters=true.